\documentstyle[epsf]{elsart}
\begin{document}
\begin{frontmatter}
\title{Estimation of distribution parameters from statistically limited
information; muons in KASCADE experiment}

\author{Tadeusz Wibig}
\address{Experimental Physics Dept., University of \L \'{o}d\'{z}, \\
Pomorska 149/153, PL-90-236 \L \'{o}d\'{z}, Poland}

\begin{abstract}
The problem of the estimation of distribution parameters in the
case of experimentally limited information is discussed.
As an example the determination of the total number
of muons in Extensive Air Showers registered in the KASCADE experiment is
studied in details.
Some methods based on other than standard maximum--likelihood approach
are examined. The advantages of the new methods are shown. The comparison
with the Artificial Neural Network approach is also given.
\end{abstract}
\end{frontmatter}

\section {Introduction}

The experimental study of a particular distribution shape is often related
to the extraction of an information from the statistically scant data.
Due to the physics of a phenomenon under study or to the experimental
situation the distribution we want to distinguish is only sampled by the
very limited number of measured occurrences. The very good example of such
situation takes place in cosmic ray ground level physics. The high energy
particle cascade initiated by the particle of energy of PeV or more reach the
sea level as a huge number of elementary particles. More that 95\% of them
are electromagnetic particles ($\mathrm{e^{-}}$, $\mathrm{e^{+}}$
or high energy gamma quanta) there are also very
few hadrons and  few percents muons. All these particles are distributed
in the plane perpendicular to the shower axis. Shapes of these
distributions in each Extensive Air Showers (EAS) contain an information
about two main points of interest in cosmic ray physics: the nature
of the primary particle and the character of the very high energy
interaction. Thus, to extract that information the distributions
have to be regain first. The electrons (positrons and gammas) are relatively
abundant so the determination of their distribution makes rather modest
problem. The muons, as it will be shown below, gives an opportunity
to look for a non-standard solutions of the distribution parameter
estimation problem with a poor statistics.

\section {Basic definitions}

Let the $f(x; \{p_m\})$ will be the ''inclusive'' distribution of some random
variable $x$. The form of $f$ is preadjusted from some theoretical assumptions
or experimental experience. There is a set of parameters $\{p_m\}$ (including
may be also a normalization constant - the $f$ need not to be normalized to
unity) to be estimated. The methods commonly used for that purposes are based
on minimization with respect to $\{p_m\}$ some distance measure
between the $n$ particular random realizations of variable $x$ and
the predictions for given values of $\{p_m\}$.
(The value of $n$ can be also a random variable, as it
will be seen in the case of muons in EAS.)
The standard procedure is to build the histogram from the
$\{x_i ; i=1,\ldots,n\}$ and use one of handbook methods to fit
the histogram using values obtained from $f(x; \{p_m\})$ by
respective integration. The most popular method is to use the
$\chi ^2$ measure.

The problem can be solved satisfactory in all the cases where the
method is applicable. In general it can be said than this happens if the
number $n$ of used random realizations of function $f$ is sufficiently
high. The meaning of ''sufficiently'' is not well define. All depends
on the way of binning. For too many
(with respect to the value of $n$) too small bins the contents of some bins
can be not enough to satisfy the $\chi ^2$ applicability conditions.
It is quite clear that if we have very few entries in a particular bin
the fluctuations of the bin contents could not be treated as gaussian
and the simple probabilistic meaning of $\chi ^2$ is
no longer valid. However the formula for calculation the distance:

\begin{equation}
u\ := \ \sum_\mathrm{histogram} {{\left( N\mathrm{e}_i - N\mathrm{c}_i
\right)^2} \over { N\mathrm{c}_i}}  ,
\label {u1}
\end{equation}

where $N\mathrm{e}_i$ is the content of $i$-th histogram bin
and $N\mathrm{c}_i$ is its
expected value for a given set of parameters, can be still treated as
a measure of a distance between the histogramed data and the prediction
for given $\{p_m\}$ remembering that the probability distribution of the
variable $u$ is not the ${\chi}^2$ one. In that sense it can be used
in minimization procedures to fit some unknown parameters for any
bin contents.

The inconvenience of statistical interpretation of the distance
defined in Eq. (\ref{u1}) can be avoid if we use the exact maximum
likelihood method. The measure used is:

\begin{equation}
u \ := \ \sum_\mathrm{histogram} \left[
\: - \ln
\left(\
p \left( N\mathrm{e}_i ;\: N\mathrm{c}_i \right)\
\right)\
 \right]   ,
\label{u2}
\end{equation}

where $p \left( N\mathrm{e}_i ;\: N\mathrm{c}_i \right)$ is the probability
that $i$-th bin of the histogram contents is $N\mathrm{e}_i$ while the
value expected for the given set of parameters $\{p_m\}$ is $N\mathrm{c}_i$.

In that paper the other possibilities of the distance measure between
histogramed data and given values of $\{p_m\}$ prediction are also
investigated.
The first proposition is the measure similar to the euclidean metric in
multidimensional space:

\begin{equation}
u \ := \  \sqrt {\sum_\mathrm{histogram}
\left( N\mathrm{e}_i - N\mathrm{c}_i \right)^2
}  ,
\label{u3}
\end{equation}

the other is:

\begin{equation}
u \ := \  \sum_\mathrm{histogram}
\left| N\mathrm{e}_i - N\mathrm{c}_i \right|
  .
\label{u4}
\end{equation}

During the binning procedure some amount of information is lost.
The problem arises in particular when the normalization constant is one
of the parameter to be estimated and the expected number of $n$ is small.
The methods which use the information about each single $x$ (or from
histograms with very small bin size, what is analogous to some extend)
establish qualitatively quite different and important
(as it will be seen) class of measures. Among them the best known are those
derived using the concept of random walk (which is, in fact, a basis of the
well known Kolmogorov--Smirnoff test).
From the set of $\{x_i\}$ something like the cumulative distribution
can be made:

\begin{equation}
D\mathrm{e}_j \ = \ \sum_{i=1}^k \Theta (x_j-x_i )   ,
\label{de}
\end{equation}

where

\begin{equation}
\Theta (x) \ = \ \left\{  \matrix { 0 & \mathrm{for} & x < 0 \cr
                                     1 & \mathrm{for} & x \geq 0 }  \right.
\label{de2}
\end{equation}

and $j$ can enumerate both: single values of $x$ or histogram channel contents.

In the same way the expected ''cumulative distribution''
(normalized with respect to the $n$) of $f(x; \{p_m\})$ obtained for
given values of parameters can be defined:

\begin{equation}
D\mathrm{c}_j \ = \ \sum_{\mathrm{j}}
\int\ \Theta (x_j-x) f(x; \{p_m\}) \d x    ,
\label{dc}
\end{equation}

where the integration is taken over the whole $x$ domain covered
experimentally.

Using both ''cumulative distributions'' the measure can be defined as:

\begin{equation}
u \ := \ \max_\mathrm{j} \left(\ \left| D\mathrm{e}_j -
D\mathrm{c}_j \right| \right)
\label{u11}
\end{equation}

Here again some modifications can be made. Two examples will be discussed
in that paper:

\begin{equation}
u \ := \
\max_\mathrm{j} \left( D\mathrm{e}_j - D\mathrm{c}_j,\ 0  \right) \ + \
\max_\mathrm{j} \left( D\mathrm{c}_j - D\mathrm{e}_j,\ 0  \right)
\label{u12}
\end{equation}

and

\begin{equation}
u \ := \
\left[\ \left( \max_\mathrm{j} \left( D\mathrm{e}_j - D\mathrm{c}_j
\ ,0 \right) \right) ^2 + \
\left( \max_\mathrm{j} \left( D\mathrm{c}_j - D\mathrm{e}_j, \ 0 \right)
\right)^2
\right]^\half  .
\label{u13}
\end{equation}

\section {Applications to the muons in KASCADE experiment}

For the statistical methods examination it is enough to test the case of
{\em ideal experiment\/}.
Particle detectors are distributed over some area and each detector responses
giving the number of muons passed through its surface. In the real situation
there is some spread of the detector signal so the results obtained in the
present work can be considered as a most optimistic approximation of the
reality.

The geometry of our {\em ideal experiment\/} studied in that paper is
exactly the geometry
of the \mbox{KASCADE} experiment (Ref.\ \cite{kascade}). This experiment is
at present about starting to collecting data and the expected quality
of these data gives a hope to make a great improvement in our knowledge
of the nature of EAS. Among the others there are 192 muon detectors of
area 3.2 $\mathrm{m}^2$ distributed over the area of 200 m $\times$
200 m.
The KASCADE experiment is dedicated to EAS induced by
primary CR particle of energies from about of $\mathrm{10^{15}}$ eV.
In that paper the low energy threshold of the experiment due to
lack of information about muon distribution from the array
detectors is investigated.

To start any {\em ideal experiment\/} a source of {\em ideal inputs\/} is
needed. The performance of the data evaluation can be tested only when
there is a set of presumed {\em inputs\/} as they are expected in real
case {\em and\/}
the respective set of {\em true answers\/} which the experiment should
give.
In our case the Monte--Carlo shower simulation code CORSIKA version 4.112
was used (Ref.\ \cite{corsika}). That program realize the complete simulation
of particle passage through the atmosphere producing the large output file
of all particles on the level of observation. For that paper the use of
particular code is not very important. The statistical
behaviour of an {\em ideal experiment\/} does not,
to some extend, depend of the physical exactness of the Monte--Carlo
calculation of EAS development. For the same reasons in the present
analysis only vertical showers are used.

The Monte--Carlo program in principle do not give the muon lateral
distribution. Some finite number of muons is nevertheless distributed
somehow over the experiment surface. Due to the random character of
physical processes in the shower development the concept of muon
lateral distribution in each individual EAS can be introduced.
It can be define as a probability distribution sampled by each
muon in Monte--Carlo simulated shower. All the information about that
distribution can be obtained, by definition, only from the finite number
of all muons in the particular shower. The accuracy of such definition for
our purposes is quite enough (the surface covered by muon detectors
allows us to measure only the fraction of percent of the muons in EAS).

The detail shape of the muon lateral distribution is of course unknown, but
some analytical formulae could be found in the literature. The particular
choice is again not very important for the present study. We used the one
which reasonably well describes the Monte--Carlo outputs:

\begin{equation}
\rho (r) \: r^2 \ = \ N_\mu \left( {r \over R_0} \right)^n \
\left( 1 + {r \over R_0} \right) ^{- (m+n)} \
{{\Gamma (n) \Gamma(m)} \over { \Gamma (n+m) }}  ,
\label{rho}
\end{equation}

where $r$ is the distance from the shower core, which position was assumed to
be known for our purposes (from electromagnetic EAS component measurement).
The normalization of the above distribution was done in the way
that the $\rho$ can be treated as a muon density at a given distance
and $N_\mu$ is the total muon number in the shower.
There are four parameter in Eq. (\ref{rho}) to be adjusted
$(N_\mu, R_0, n \ \mathrm{and} \ m)$.
It has been checked that the value of $m$ can be fixed without loosing the
quality of the {\em ideal inputs\/} description. The value of $m = 1.7$ is used
hereafter. This was possible mainly because of the fact that the distances
from the shower core to detectors were limited to about $<$ 300 m
(the simulated showers were uniformly distributed within about 50 m
form the center of the KASCADE array).

For each simulated shower the formula in Eq. (\ref{rho}) was fitted
with the help of standard methods
to all muons in the distance range from 10 m to 300 m. The exactness of
such fits is given below.
The obtained parameters $N_\mu$, $R_0$ and $n$ will be hereafter called
{\em true\/} parameters of the particular simulated shower.

In the next step the muons from the simulated shower are sampled by the
net of our {\em ideal\/} muon detectors producing the output signal of  the
{\em ideal experiment\/}. Detectors can be treated as a relatively small bins
in the histogram with respect to the distance from the shower core.

The general problem is how the information about {\em true\/} values of the
individual shower muon distribution parameters $(N_\mu, R_0, n)$
can be derived from the 192 histogram channel contents.

First method tested was the standard $\chi ^2$ method. The
histogram (detector responses) was rebind. Finally, after some tests,
the constant bin limits
have been chosen:
(10, 20, 30, 50, 70, 100, 130, 170 and 200 m). The sum in
Eq. (\ref{u1}) was performed only for the bins with the contents
not less than 5.
That method will be hereafter called H$\chi$1.

Without assumption about the minimal bin contents the Eq. (\ref{u1})
was also used. This method will be called $H\chi$2.

Next tested possibility is associated with the Eq. (\ref{u2})
The assumption that all the muons are uncorrelated (for one particular EAS)
leads to the poissonian distribution of the fluctuations in
each histogram bin:

\begin{equation}
p \left( n ;\: \overline {n} \right) \ = \
{ \overline {n} ^n \over {n !}}
\ {\e}^{- \overline{n}}  .
\label{poiss}
\end{equation}

Using that directly with Eq. (\ref{u2}) the method called HL
is defined.

The equations Eqs. (\ref{u3}) and (\ref{u4}) applied to the histogramed
detector outputs leads to the methods labeled HE, HA
respectively.

The Kolmogorov--Smirnoff like method was also checked for the rebind
histogram. The Eq. (\ref{u11}) where $j$ indexes 10 big bins
lead to the method called HK.

The main aim of that work is to look for the best estimation method. It
was foreseen that using directly each detector information the results
should be improved. First the standard measure (Eq. (\ref{u1})) was checked.
( With $N\mathrm{e}_i$ replaced in that case by the $i$-th detector
response and $N\mathrm{c}_i$ by its expected value. )

The assumption of minimum hits in particular detector was of course rejected.
That method will be called D$\chi$.

With analogy to the method introduced to the big bin histogram the measures
defined by Eqs. (\ref{u3}) and (\ref{u4}) were used leading to the methods
denoted by DE and DA respectively:

\begin{equation}
u \ := \  \sqrt {\sum_\mathrm{detectors}
\left( n_i - s\ \rho(r_i) \right)^2
}  ,
\label{u31}
\end{equation}

\begin{equation}
u \ := \  \sum_\mathrm{detectors}
\left| n_i - s\ \rho(r_i) \right|
  ,
\label{u32}
\end{equation}

where $s$ is a detector area, $r_i$ its distance from the shower core
and $n_i$ detector response.

The maximum likelihood methods are the standard tools in the estimation theory.
In the present work that possibility was investigated too. In general, it is
based on the
knowledge of the shape of fluctuations with respect to the mean of the
muon number passing the given detector surface at fixed distance form the
shower core. The detail study of the Monte--Carlo simulated showers
allows us to state that they are wider than poissonian (Ref.\ \cite{twr}).
They can be well
approximated by the negative binomial distribution:

\begin{equation}
p(n;\: \overline{n}, \gamma )\ = \
{{\overline{n} / \gamma + n - 1} \atopwithdelims() {n}}
\ \left[\: {{1/\gamma} \over {1+1/\gamma}}\right]^{\overline{n} \over \gamma}
\ \left[ {{1/\gamma} \over {1+1/\gamma}} \right]^n
\label{u9}
\end{equation}

with $\gamma = 1.0$. With that assumption the maximum likelihood
measure can be defined (and than minimize):

\begin{equation}
u \ := \ \sum_\mathrm{detectors} \left[ {- \ln \left( \:
p \left( n_i ;\: s \: \rho (r_i),\: 1.0 \right) \:\right)}\: \right]  .
\label{u10}
\end{equation}

That method will be hereafter called DP1.

To see if the particular negative binomial shape of the
fluctuations produce an important effect the measure defined
by Eq. (\ref{u2}) was used with the poissonian fluctuations
(Eq. (\ref{poiss})). The results of such a calculations is denoted by
DP2.

The Eq. (\ref{u11}) is of particular interest for the individual detector
data. The minimization of that measure leads to the method called hereafter
DD.
The modifications described in Eqs. (\ref{u12}) and (\ref{u13}) were
also tested. First named DD1 and the second DD2.

The last method results of which are presented in this paper is based on
completely different way of solving the estimation problem. This is the
Artificial Neural Network (ANN) method. The much more detailed
discussion of it will be given elsewhere (Ref.\ \cite{ann}).

For the present analysis the
neural network contains 192 input nodes which one steering with an
integer response of one detector in the array. There are two hidden
layers with successively decreasing numbers of neurons and finally
one output node. The output signal of the network is related to one
of the parameters $(N_\mu)$ of the muon lateral distribution
(Eq. \ref{rho}). The number of weights of the network to be adjusted
is very large so for the network training about hundreds of thousands
showers were used. A special technic was developed to achieve this. Details
are not very important for the present work. Finally we obtained the
network trained with the proton induced vertical showers in the energy
range of interest. The ANN method results presented here
are given just to compare them with the others and should be treated
as, more or less, preliminary.

\section{Results concerning the total number of muon estimation in the
KASCADE experiment}

The parameter of the great importance for the EAS study is a
total muon number in the shower $N_\mu$.

To perform the minimization of the respective $u$ measure
the standard MINUIT from the CERN library was used (Ref.\ \cite{minuit}).
It is obvious that in some cases due to fluctuations in the input
data the minimum of some distance measure can not be reached within
reasonable limits of the parameters in Eq. (\ref{rho}). The
limits used were 10 $\div$ 800 m for $R_0$, 0.05 $\div$ 1.99 for $n$.
The minimization was started with two parameters fixed: $R_0$ at 300 m
and $n$ at 1.43 (the values about the mean for $10^{15}$ eV) and the
minimization was performed with only one parameter to minimize ($N_\mu$).
Then the $n$ parameter was liberated and the minimization stared again from
the just found values. If possible, next, the $R_0$ parameter was
fitted also.

Simulated showers which produced all 0 in our {\em ideal\/} array response
matrix are of course lost.

All methods were examined for three different primary cosmic ray particle
energies.
As it has been said the KASCADE experiment is designed to study the
cosmic ray primary spectrum in the very interesting and unexplored up to
now region
starting at about $\mathrm{10^{15}}$ eV. Just below that energy its the
end of the area covered by the recent top of the atmosphere experiments data.
The correspondence of the direct (balloon--borne) and indirect (EAS) methods
is one of the important point of that experiment so the determination of
EAS parameters at the lower part of the shower size spectrum is very important.

\begin{figure}
\epsfxsize 250pt
\centerline{\epsfbox[0 0 750 500]{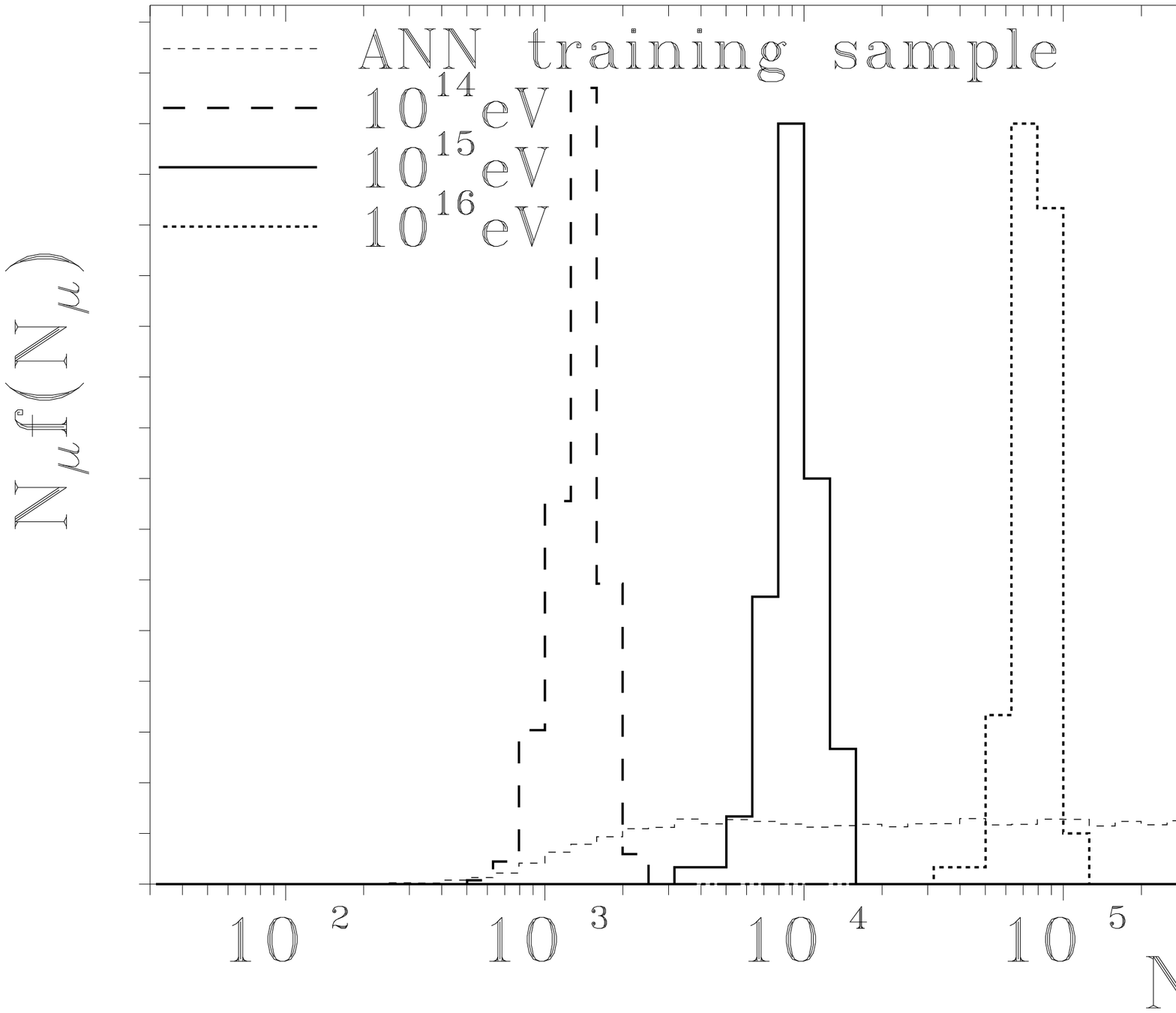}}
\caption{
Distribution of {\em true\/} $N_\mu$ values for showers of different primary
proton energies $\mathrm{10^{14}}$, $\mathrm{10^{15}}$ and
$\mathrm{10^{16}}$ eV.
The thin dashed line shows the muon shower sizes distribution of the
artificial neural network training sample.}
\label{meanm}
\end{figure}

The distributions of {\em true\/} total number of muons in showers of the
energies of interest are given in Fig.\ \ref{meanm}.

\begin{figure}
\vspace{1cm}
\epsfxsize 250pt
\centerline{\epsfbox[0 0 750 500]{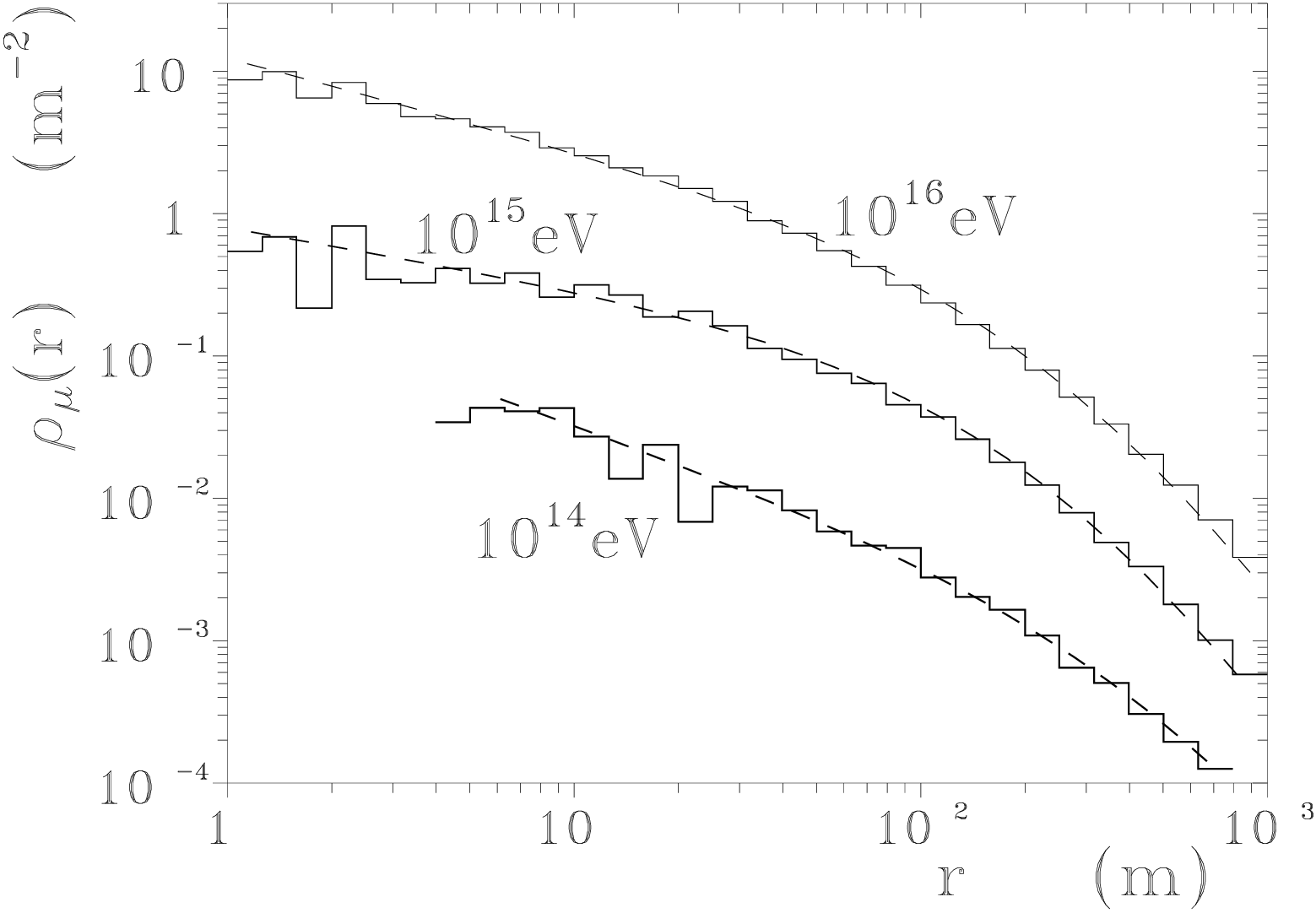}}
\caption{Examples of the muon lateral distributions for individual proton
showers of different energies (histograms).
The curves represents the {\em true\/} muon lateral distributions of that
particular showers.}
\label{mud}
\end{figure}

To better recognize the problem the typical muon lateral density
distributions of the showers initiated by the
primary vertical proton of the energies from $\mathrm{10^{14}}$ eV to
$\mathrm{10^{16}}$ eV are given in Fig.\ \ref{mud}. The {\em true\/}
muon distributions (fits of form given by Eq. (\ref{rho}) to the
histograms) are also presented.

\begin{figure}
\vspace{1cm}
\epsfxsize 300pt
\centerline{\epsfbox[0 0 397 113]{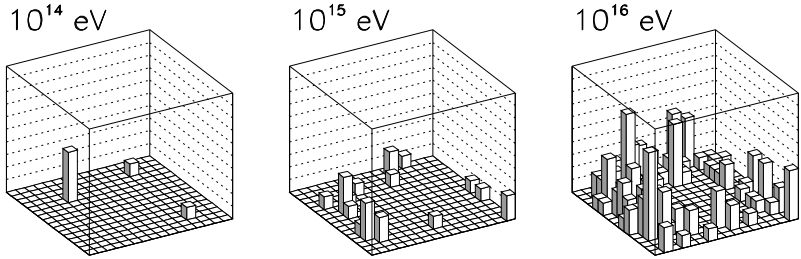}}
\caption{Examples of the muon response array for showers in Fig. 2.}
\label{expm}
\end{figure}

The spread of the points in the Fig.\ \ref{mud} is a result of the physical
fluctuations in the shower development. To make this figure all the muons
in the showers were used. The real problem arise when distributions like that
are sampled with the net of detectors which covered only few \% of the
shower front area.
The particular detector array responses for the showers which muon lateral
distributions were given in Fig.\ \ref{mud} are presented
in Fig.\ \ref{expm}.

It can be seen that the expected numbers of muons seen in detectors is
rather small as well as the fraction of fired detectors both for
$\mathrm{10^{14}}$ eV and even for $\mathrm{10^{15}}$ eV energies.
The situation for the energy $\mathrm{10^{16}}$ eV looks better.
It is interesting that even for the worst case some of discussed
methods could give the reasonable estimation
of the total muon contents.

The value of $N_\mu$ obtained for each shower in the {\em ideal experiment\/}
using different methods described above has to be compared with its
{\em true\/} value.
The collection of figures presented below shows the spread of the calculated
values about the {\em true\/} ones.
The number of showers used for each test
was about 3000 and the histograms were normalized in such a way that the area
below each one is related to the fraction of cases were given
method was used.
The shapes of the histograms give a
self--explainable answer to the question about the exactness of each of the
methods. The labels gives the references to the particular method for which
the histograms were obtained.

\begin{figure}
\centerline{\epsfbox[0 0 340 354]{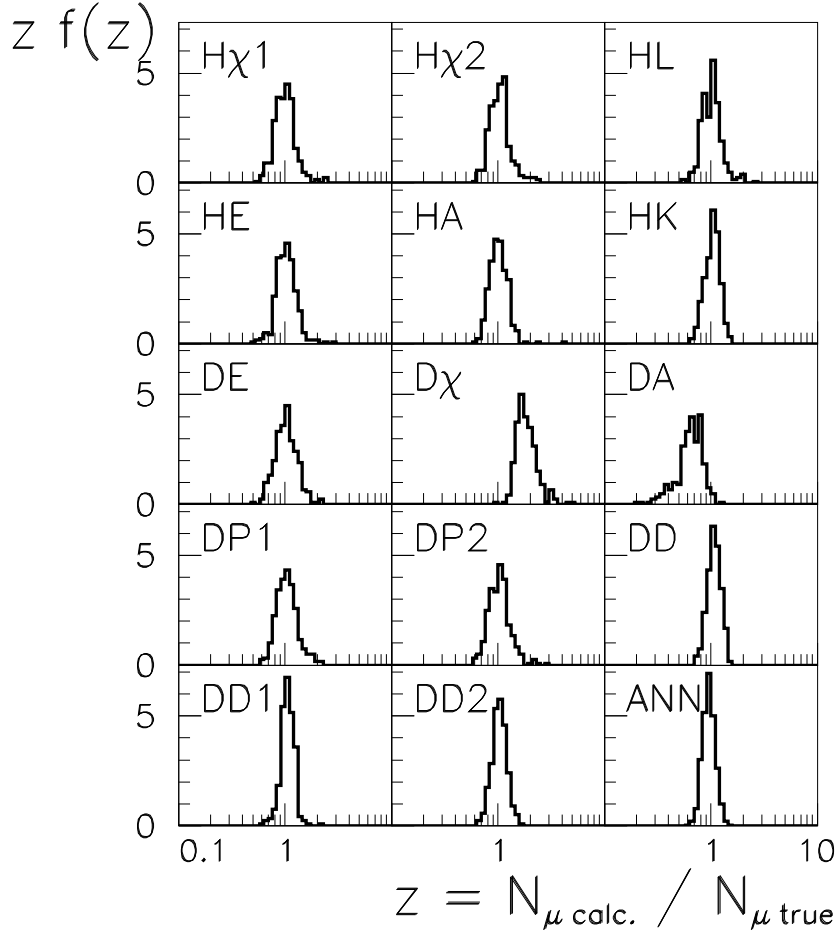}}
\caption{
The spread of the estimated $N_\mu$ value for each method discussed in the
paper for the vertical $\mathrm{10^{16}}$ eV proton induced showers.
}
\label{1e7}
\end{figure}

\begin{figure}
\centerline{\epsfbox[0 0 340 354]{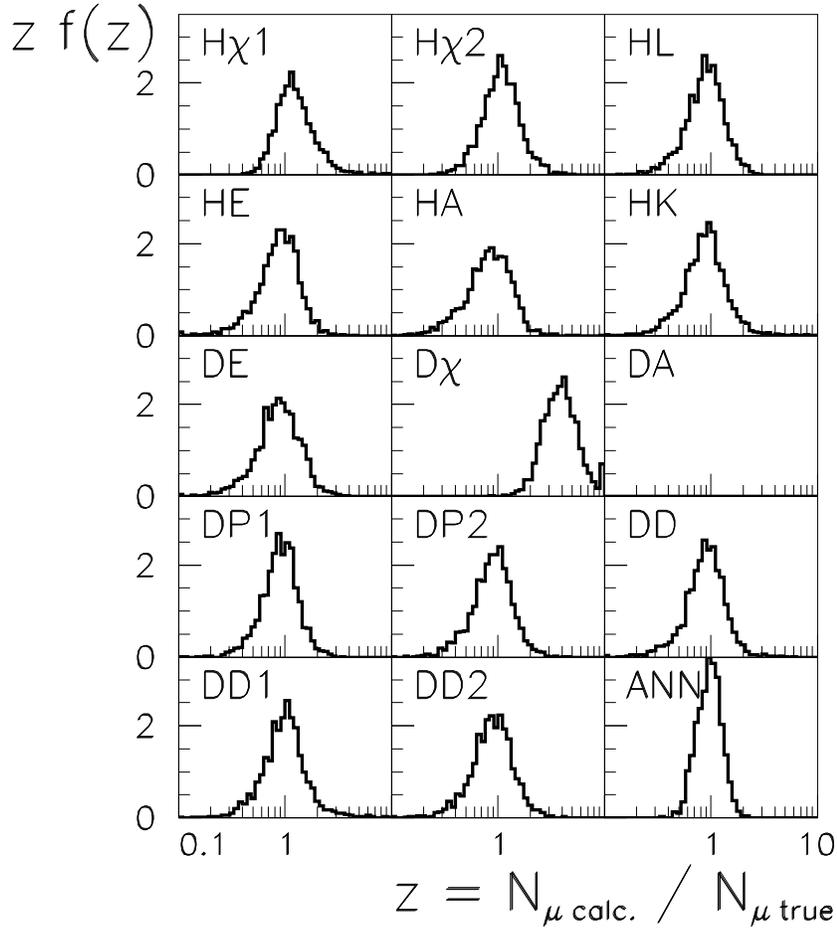}}
\caption{
The spread of the estimated $N_\mu$ value for the $\mathrm{10^{15}}$ eV
proton induced showers.
}
\label{1e6}
\end{figure}

\begin{figure}
\centerline{\epsfbox[0 0 340 354]{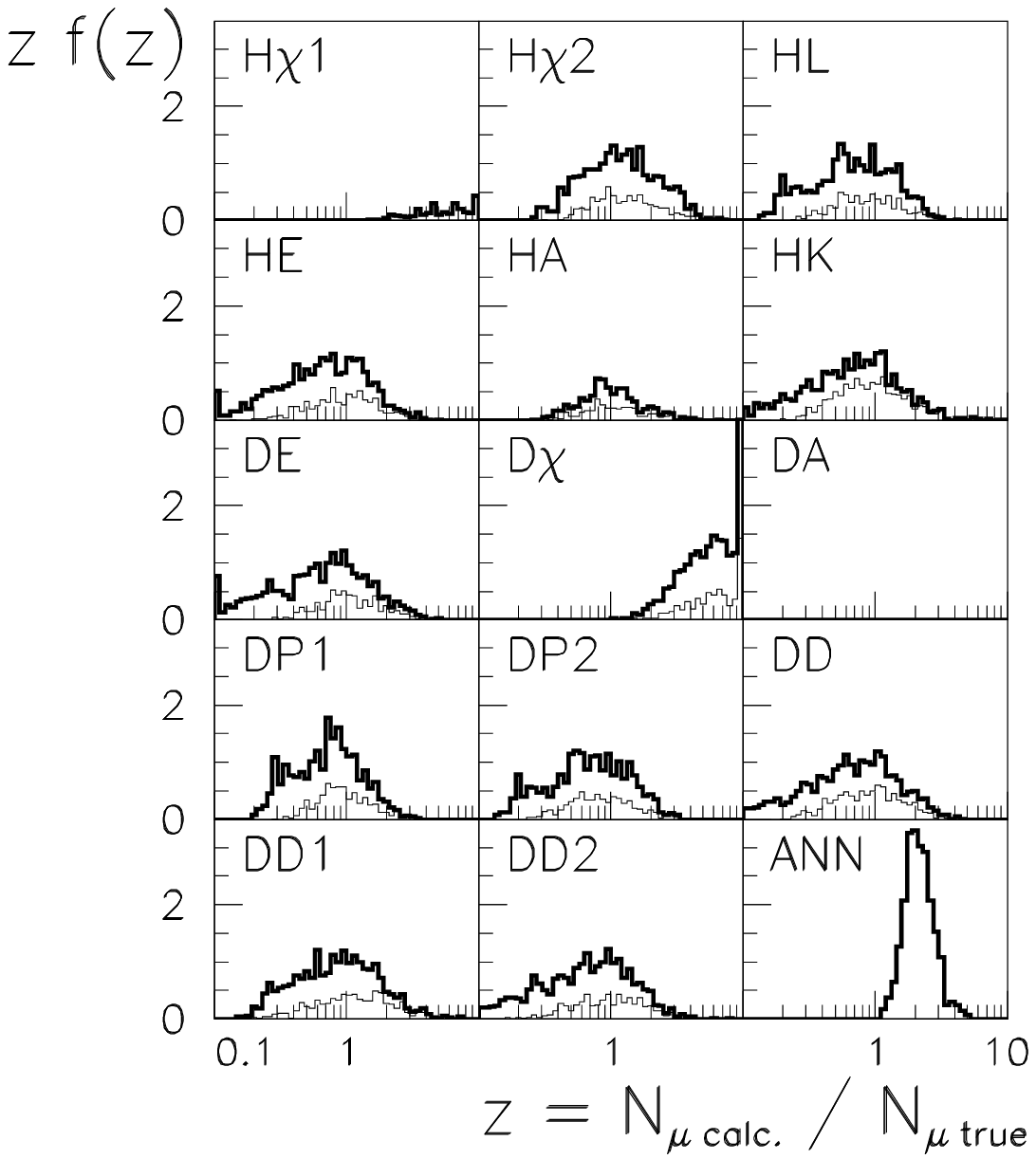}}
\caption{
The spread of the estimated $N_\mu$ value for the $\mathrm{10^{14}}$ eV
proton induced showers.
The thin lines show results for events for which
at least two parameter fit has been successfully applied.
}
\label{1e5}
\end{figure}

In Fig.\ \ref{1e7} there are shown results for $\mathrm{10^{16}}$ eV.
Figs.\ {\ref{1e6}} and {\ref{1e5}} present quality of the methods for showers
with energies smaller of one and two orders of magnitude respectively.

The more quantitative results for different methods comparison are given
in the Tables.
In the $\mathrm{2^{nd}}$ column mean deviations of logarithm of the
calculated $N_\mu$ from its {\em true\/} value are shown.
In the $\mathrm{3^{rd}}$ column the accuracy of each method
defined as a $\sigma$ of
the $\left( \log_{10}N\mathrm{c}_\mu / N\mathrm{e}_\mu \right)$
distribution (as they are presented in Figs. \ref{1e7} -- \ref{1e5}).
The applicability of each method defined as a
fraction of showers for which particular method was successfully used is
shown in the $\mathrm{4^{th}}$ columns. In the $\mathrm{5^{th}}$ column
in the Table \ref{table1e5} the fraction of events for which at least two
parameter fit was possible is given.

\begin{table}
 \begin{center}
\begin {tabular} {|l|r|r|c|}
\hline
method & $\left< \log_{10} \right>$ & $ \sigma_{\log} $ & applicability\cr
\hline
H$\chi$1&  0.017 & 0.099 & 1.000\cr
H$\chi$2&  0.038 & 0.100 & 1.000\cr
HL      &  0.027 & 0.097 & 0.993\cr
HE      & 0.030 & 0.097 & 1.000 \cr
HA      & 0.026 & 0.088 & 1.000 \cr
HK      & 0.032 & 0.070 & 1.000 \cr
DE      & 0.029 & 0.103 & 1.000 \cr
D$\chi$ & 0.287 & 0.097 & 1.000 \cr
DA      & -0.180& 0.125 & 0.997 \cr
DP1     & 0.037 & 0.100 & 1.000 \cr
DP2     & 0.032 & 0.102 & 0.990 \cr
DD      & 0.048 & 0.062 & 1.000 \cr
DD1     & 0.042 & 0.066 & 1.000 \cr
DD2     & 0.031 & 0.070 & 1.000 \cr
ANN     & -0.021 & 0.063 & 1.000 \cr
\hline
\end{tabular}

\end{center}
\caption{Detailed results for $\mathrm{10^{16}}$ eV  proton shower sample.}
\label{table1e7}
\end{table}
\begin{table}
\begin{center}
\begin {tabular} {|l|r|r|c|}
\hline
method & $\left< \log_{10} \right>$ & $ \sigma_{\log} $ & applicability \cr
\hline
H$\chi$1&  0.117 & 0.188 & 0.814 \cr
H$\chi$2&  0.048 & 0.186 & 0.987 \cr
HL      & -0.043 & 0.188 & 0.987 \cr
HE      & -0.049 & 0.207 & 0.987 \cr
HA      & -0.072 & 0.218 & 0.932 \cr
HK      & -0.040 & 0.213 & 0.987 \cr
DE      & -0.050 & 0.214 & 0.987 \cr
D$\chi$ &  0.612 & 0.178 & 0.987 \cr
DA      &        &       &       \cr
DP1     & -0.032 & 0.175 & 0.984 \cr
DP2     & -0.039 & 0.194 & 0.978 \cr
DD      & -0.040 & 0.195 & 0.987 \cr
DD1     &  0.011 & 0.212 & 0.987 \cr
DD2     & -0.041 & 0.202 & 0.987 \cr
ANN     & -0.019 & 0.116 & 0.987 \cr
\hline
\end{tabular}
\end{center}
\caption{Detailed results for $\mathrm{10^{15}}$ eV  proton shower sample.}
\label{table1e6}
\end{table}

\begin{table}
\begin{center}
\begin {tabular} {|l|r|r|c|c|}
\hline
method & $\left< \log_{10} \right>$ & $ \sigma_{\log} $ & applicability
& at least 2 par. fit \cr
\hline
H$\chi$1 & 0.752  & 0.245 & 0.12 & 0.110\cr
H$\chi$2 & 0.087  & 0.283 & 0.89 & 0.269\cr
HL       & -0.160 & 0.304 & 0.89 & 0.261\cr
HE       & -0.175 & 0.344 & 0.89 & 0.278\cr
HA       &  0.011 & 0.236 & 0.32 & 0.142\cr
HK       & -0.162 & 0.346 & 0.89 & 0.415\cr
DE       & -0.194 & 0.374 & 0.89 & 0.246\cr
D$\chi$  &  0.808 & 0.255 & 0.89 & 0.242\cr
DA       &        &       &      & \cr
DP1      & -0.130 & 0.250 & 0.88 & 0.262\cr
DP2      & -0.158 & 0.299 & 0.86 & 0.250\cr
DD       & -0.165 & 0.341 & 0.89 & 0.338\cr
DD1      & -0.081 & 0.311 & 0.89 & 0.309\cr
DD2      & -0.175 & 0.344 & 0.89 & 0.260\cr
ANN      &  0.331 & 0.112 & 0.89 & \cr
\hline
\end{tabular}
\end{center}
\caption{Detailed results for $\mathrm{10^{14}}$ eV  proton shower sample.}
\label{table1e5}
\end{table}

For the neural network the total muon number $N_\mu$ was only one
parameter to study and a value of it was created for each shower
except ''all 0'' showers which were not used in the present analysis.

\section{Discussion of the results concerning the total number of
muon estimation in the KASCADE experiment}

Staring from the highest energy used for present examinations,
$\mathrm{10^{16}}$ eV, some interesting statements can be given
about the efficiency
of different methods for relatively muon rich showers
($N_\mu \sim \mathrm{10^5}$) where in every case at least about
50 detectors were hit.
As to the methods based on a histogramed data it is seen that all of them
lead to very similar results. The reason for that is clear. The statistical
significance of the information stored in the big bin histogram is so
large that the particular differences between different measures used in
minimization procedures has no great effect on the results. The situation is a little different
when using the raw detector data. The biases are seen for two measures:
D$\chi$ and DA.
Among the others it is clear that the two using the
''cumulative distribution'' metrics leads to the smallest spread of the
calculated $N_\mu$ value with respect to the {\em true\/} one.

Comparison of the best minimization methods (e.g. DD or DD1) with
the artificial neural network approach leads to the
conclusion that no significant difference is seen. This can be treated as an
evidence for the assumption that the exactness of the $N_\mu$ value estimation
reaches its limit allowed by the physical fluctuations of the shower
development. If two such different ways of the data evaluation shows as good
agreement that with the high level of confidence one can postulate that it is
a real limit, and there is no other way to get more information (about the
parameter under the study) from that data.

Going down with the energy to $\mathrm{10^{15}}$ eV, it is clearly seen
in Fig.\ \ref{1e6} and the Tables
that the quality of all methods decreases.
When the number of registered muons
is really a few per event some methods fail in general but, what is
surprising, some are still useful.

The differences of using
the histogramed data measures appear. Some of them become unapplicable in a
fraction of events for which the others are still useful. The larger
differences arise for the raw data. The method DA for such showers
can not be used any more for showers of that energy. The bias seen
for D$\chi$ method become so great that is usefulness has to be
also questionable.
It is very interesting to note the difference between the two ''maximum
likelihood'' method DP1 and DP2. The assumption about the poissonian
fluctuations
of the number of muons in single detector leads to enlargement of the error
of the DP2 method in comparison with DP1. The reason for that is in the events
in which the relatively big fluctuation on one detector appears.
Its probability calculated from the poissonian distribution is
much smaller that it is in real shower situation approximated by the
negative binomial density fluctuation distribution. The significance in the
minimization procedure of that
particular detector increases when the mean muon density
at a detector site is small.

The shapes of the most of thick histograms in Fig. \ref{1e5} is not gaussian.
This is simple a result of the details of the minimization technic used.
The initial values of the $R_0$ and $n$ parameters were taken to be the average
values for $\mathrm{10^{15}}$ eV showers. Thus, for a very small showers
when the information about the muon distribution is very limited those
parameters sometimes could not be adjusted further. Because the small
showers are in average distributed wider in shower front plane the total
muon number obtained from the fit of only one parameter has to be shifted
to the smaller values. In the Fig. \ref{1e5} there are also shown by the
thin line histograms the results for the showers for which at lest one
of the distribution shape parameters could be adjusted. Those histograms are
much more gaussian.

Among the minimization methods which works well for those small showers
it is hard to chose the one significantly better than the others specially when
trying to compare events with at lest two parameters fitted.

The method based on the ''cumulative distribution'' measures (both: using the
histogramed data - HK, and that using raw data - DD, DD1 ) could be used
for at least two parameter fit in the largest fraction of events. The
width of the $\left( \log_{10}N\mathrm{c}_\mu / N\mathrm{e}_\mu \right)$
distribution is the smallest for the DP1 method. ( Tab.\ \ref{table1e5} )

Some advantages of ANN appear much clearly.
However the ANN method for that small showers introduces
systematical bias toward higher $N_\mu$ values. The reason of
that is clear. In Fig.\ \ref{meanm} there is presented {\em true\/} shower
muon size spectrum for training sample. The network was trained only with
the showers for which at least three detectors were hit while most of the
showers of $\mathrm{10^{14}}$ eV fire not more than five detectors. The
value of the bias is correlated with the number of fired detectors (Ref.\
\cite{ann}) so it
can be removed ''by hand'' from the ANN output. There is
another way of removing that bias. It is expected that if there will be
used some real physical trigger for the ANN shower training
sample and the same trigger will be used later in the ideal experiment
tests the bias will be automatically removed. In such a case even the
all zero events should produce a reasonable (to some extend) ANN answer.

\section{Conclusions}

Final conclusions of the present work can be formed in a few statements:

\begin {itemize}

\item For large showers the most promising are the methods based on
the minimization procedure of the ''cumulative distributions''
inspired measures using individual detector data (DD and DD1) .
\vspace{5mm}

\item For small showers some of the methods using
the ''cumulative distributions'' are slightly better than the others.
\vspace{5mm}

\item The knowledge of the fluctuation on each particular detectors
allows one to use right the maximum likelihood method. That knowledge improves
to some extend the accuracy of the parameter estimation.
\vspace{5mm}

\item For the studied in the present paper {\em ideal experiment\/}
of the geometry of \mbox{KASCADE}
(its array muon part) the best developed methods allows one to go down with the
muon sizes of analyzed showers to about few thousands per shower (below
$\mathrm{10^{15}}$ eV).
\vspace{5mm}

\item The artificial neural network approach gives a hope that the limit
can be shifted down to $\mathrm{10^{14}}$ eV. However that methods is strongly
effected by the exactness of the simulations used for the network training.

\end{itemize}


\begin{thebibliography}{9}

\bibitem{kascade}
P. Doll et al., Kernforschungszentrum Karlsruhe Report KFK 4686 (1989);
K. Bekk et al., Proc. Inter. Cosmic Ray Conf. Calgary {\bf 4} (1993), 674.

\bibitem{corsika}
J. N. Capdevielle et al., Kernforschungszentrum Karlsruhe Report
KFK 4998 (1992).

\bibitem{twr}
T. Wibig,{\em Comparison of some properties of Extensive Air Showers
simulated by CORSIKA with existing experimental data\/},
Kernforschungszentrum Karlsruhe Internal Report, 1990 (unpublished).

\bibitem{ann}
T. Wibig, in preparation.

\bibitem{minuit}
MINUIT, CERN Program Library Long Writeup D506 ,CERN (1992).

\end{thebibliography}
\end{document}